# PMC text mining subset in BioC: 2.3 million full text articles and growing

Donald C. Comeau, Chih-Hsuan Wei, Rezarta Islamaj Doğan and Zhiyong Lu

National Center for Biotechnology Information, U.S. Library of Medicine, National Institutes of Health, Bethesda, MD, USA

Interest in full text mining biomedical research articles is growing. NCBI provides the PMC Open Access and Author Manuscript sets of articles which are available for text mining. We have made all of these articles available in BioC, an XML and JSON format which is convenient for sharing text, annotations, and relations. These articles are available both via ftp for bulk download and via a Web API for updates or more focused collection. Availability: https://www.ncbi.nlm.nih.gov/research/bionlp/APIs/BioC-PMC/

Keywords

PMC, full text, text mining, BioC

## Introduction

Text mining of the full text biomedical research literature is growing in importance. There are an increasing number of corpora and studies [1-10]. Meanwhile, more full text is becoming available for text mining. The National Center for Biotechnology Information (NCBI) resource, PubMed Central® (PMC), is a collection of biomedical research literature available to read on the web. The PMC Open Access Subset is a well-known portion of the PMC articles under a Creative Commons or similar license that allows more liberal reuse than traditional copyright.[1] Less well-known is the Author Manuscript Collection. These articles have been made available in compliance with the NIH Public Access Policy or similar policies of other funders.[2] Figure 1 shows that the proportion of PMC articles available in the Open Access Subset and the Author Manuscript Collection is steadily growing. This is very exciting for the biomedical text mining community.

---

[1] https://www.ncbi.nlm.nih.gov/pmc/tools/openftlist/
[2] https://www.ncbi.nlm.nih.gov/pmc/about/mscollection/

# Growth of Text Mining Collection in PMC

As of 2017: 4,758,920 in PMC. 1,880,977 in Open Access (OA), and 577,043 Author Manuscript (AU).

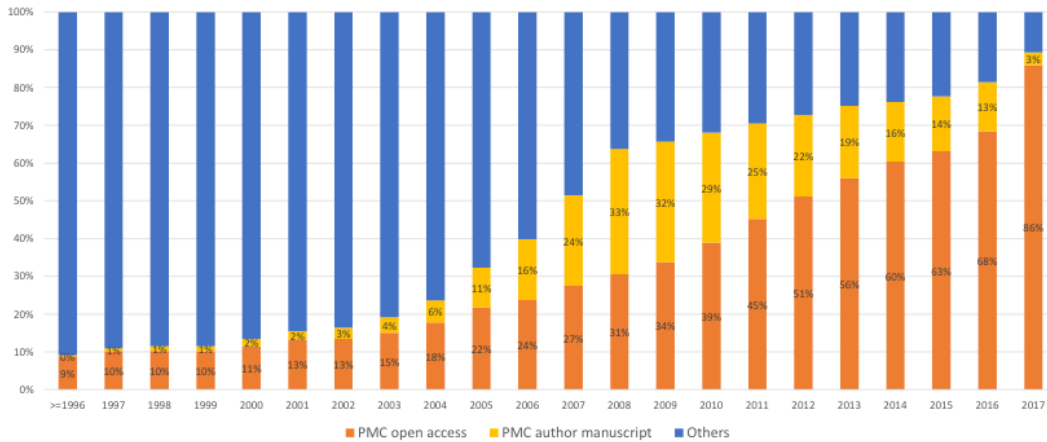

*Figure 1 Proportion of PMC articles available in the Open Access Subset and the Author Manuscript Collection. (The Others portion of the 2017 bar is smaller than might be expected because of the one-year embargo on some articles.)*

PMC articles are encoding using the Journal Article Tag Suite (JATS).[3] This is a powerful, complicated system that tracks all available meta-information and allows articles to be displayed in a manner very similar to the way they appear in a printed journal. One sign of this flexible system is that 277 XML elements are defined. But JATS is not designed to aid text mining. The text appears at different levels and in various structures. There is markup, that while potentially informative, is also another complication to be addressed.

The BioCreative community has previously developed BioC, which is a simple, straightforward data structure for dealing with text, annotations on that text, and relations between those annotations [11]. Other methods have also been developed for sharing text and annotations such as: PubAnnotation[4] [12], which is a public repository aiming to collect all annotations on PubMed® articles, with purpose of combining and disseminating them to the community for further use and reuse, and BRAT[5], which is a web-based tool for text annotations, that particularly facilitates the structural annotation to denote relations between entities.

Unlike some other formats, BioC easily handles both PubMed abstracts and PMC full text articles. Serializations are now available for both JSON and XML. JSON is very popular because the output of JSON parsers is more convenient to use than a traditional XML DOM. Even with

```
<p>R<sc>ecently published studies</sc> in peer-reviewed journals<sup><xref rid="B1" ref-
type="bibr">1–4</xref></sup> and high-profile articles in the <italic>New Yorker</italic>,<sup><xref
rid="B5" ref-type="bibr">5</xref></sup><italic>New York Times</italic>,<sup><xref rid="B6" ref-
type="bibr">6</xref></sup> and <italic>Wall Street Journal</italic>,<sup><xref rid="B7" ref-
type="bibr">7</xref></sup> have rekindled professional and public interest in the therapeutic use of
psychedelic drugs. It is easy to understand the enthusiasm. The magazine and newspaper articles include
accounts of patients with profound depression, demoralization associated with terminal illness, and anxiety
related to post-traumatic stress disorder (PTSD), who experienced remarkable improvements, including some
who had previously considered suicide.</p>
```

*Figure 2 Sample portion of a PMC XML document*

additions for annotations and relations, BioC XML only uses 15 elements. Even more valuable, for XML there are libraries that populate and preserve native data structures in a number of different languages (C++, Go, Java, Python, Perl, Ruby) [13, 14]. Because of these libraries, text, annotations and relations can be painless shared between multiple tools. These libraries provide even more of an advantage for BioC XML over PMC XML.

Figure 2 shows a short sample of PMC XML. There are a number of tags, such as <xref>, that while meaningful, will not be helpful for most text mining tasks. Figure 3 shows the same portion of the document in BioC XML. Other than the information that it is a paragraph, and its location in the document, the only information is the text of the passage. Figure 4 shows the same information in BioC JSON. Here one sees the fields that can be used to hold annotation and relation information. But they are empty at the moment. Finally, Figure 5 shows the data portion of the C++ class that would hold this information. If one uses a BioC XML library, this is the only necessary view of the data.

## Methods

PMC Open Access articles have been available in BioC XML through FTP since 2014 [5]. This work makes three useful additions:

- Web API for individual or small sets of articles
- BioC JSON as an alternative to XML
- Author Manuscripts adds more than 400,000 additional articles

```
<passage>
  <infon key="type">paragraph</infon>
  <offset>1853</offset>
  <text>Recently published studies in peer-reviewed journals and high-profile articles in the New
Yorker,New York Times, and Wall Street Journal, have rekindled professional and public interest in the
therapeutic use of psychedelic drugs. It is easy to understand the enthusiasm. The magazine and newspaper
articles include accounts of patients with profound depression, demoralization associated with terminal
illness, and anxiety related to post-traumatic stress disorder (PTSD), who experienced remarkable
improvements, including some who had previously considered suicide.</text>
  </passage>
```

*Figure 3 Sample portion of a BioC XML document*

```
    {
        "text": "Recently published studies in peer-reviewed journals and high-profile articles in the New
    Yorker,New York Times, and Wall Street Journal, have rekindled professional and public interest in the
    therapeutic use of psychedelic drugs. It is easy to understand the enthusiasm. The magazine and newspaper
    articles include accounts of patients with profound depression, demoralization associated with terminal
    illness, and anxiety related to post-traumatic stress disorder (PTSD), who experienced remarkable
    improvements, including some who had previously considered suicide.",
        "offset": 1853,
        "relations": [],
        "infons": {
          "type": "paragraph"
        },
        "sentences": [],
        "annotations": []
    },
```

*Figure 4 Sample portion of a BioC JSON document.*

While accessing the text of a paper is much simpler, BioC still preserves a lot of information about a PMC article. Section title types identify the depth the of the following section, subsection, or subsubsection. Of particular interest, figure captions and table captions are preserved and identified. These sections often contain some of the most important information in the article.

A research article is typically arranged in an outline format, as in Figure 6. This means that in addition to having multiple sections, there are nested subsections, subsubsections, and so on. A BioC document contains a linear sequence of passages. If one just wants to process the text and does not care about the structure, then simply processing the passages is straightforward. However, if the structure is important, it can be recovered. The depth of a section or subsection title is indicated by the passage type: title_1, title_2, etc. If the nested hierarchical structure is important, it can be recovered by tracking these subsection titles. This linear structure, with the hierarchical passage types, is show in Table 1. The text of each passage is clear, and the passage types can be processed or ignored, as needed.

```
class Passage {
public:
  std::map<string,string> infons;
  int offset;
  string text;
  vector<Sentence> sentences;
  vector<Annotation> annotations;
  vector<Relation> relations;
};
```

*Figure 5 C++ class storing a portion of a document*

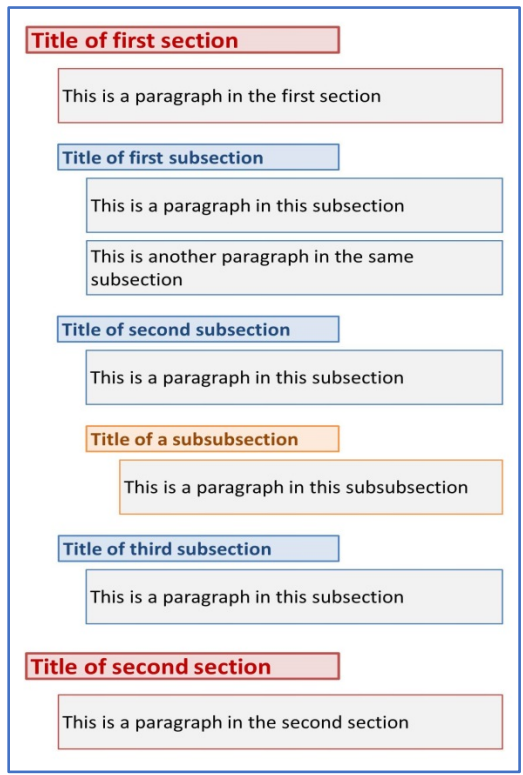

*Figure 6 Thumbnail web view of an article.*

| Passage type | Text of passage |
| --- | --- |
| title_1 | Title of first section |
| paragraph | This is a paragraph in the first section |
| title_2 | Title of first subsection |
| paragraph | This is a paragraph in this subsection |
| paragraph | This is another paragraph in the same subsection |
| title_2 | Title of second subsection |
| paragraph | This is a paragraph in this subsection |
| title_3 | Title of a subsubsection |
| paragraph | This is a paragraph in this subsubsection |
| title_2 | Title of third subsection |
| paragraph | This is a paragraph in this subsection |
| title_1 | Title of second section |
| paragraph | This is a paragraph in the second section |

*Table 1: The linear structure of **Figure** represented in BioC passages. The types allow recovery of the nested structure if desired.*

When PMC Open Access articles in BioC were first made available in BioC, they were only available as large FTP files. This is convenient for obtaining a large number of articles. For example, when first starting to work with BioC PMC. But it is not convenient for a subset of articles of interest. As a complement to FTP access, a RESTful web service is now available. It is now possible to download exactly the articles needed. This is much more convenient for small collections or for updating an existing large collection.

An original PMC article is encoded in UTF-8 Unicode. But sometimes, and with some tools, it is more convenient to process articles in ASCII. Articles are available using a Unicode to ASCII translation that we have found useful and convenient. If you would like to use your own translation, you are welcome to download in Unicode and perform your own translation.

In addition to XML, articles are now available in JSON. The JSON data structure is the same BioC internal data structures as before, except stored in JSON. That means objects are typically available as dictionaries, as is expected for JSON. The output of JSON parsers is usually convenient to work with, unlike an XML DOM. For this reason, there are, with one exception,[6] no libraries to convert between JSON and internal classes.

As mentioned above, PMC Open Access and Author Manuscript articles are available in BioC via a Web API. This API is described at https://www.ncbi.nlm.nih.gov/research/bionlp/APIs/BioC-PMC/. The URLs follow this template:

```
https://www.ncbi.nlm.nih.gov/research/bionlp/RESTful/pmcoa.cgi/BioC_[format]/[ID]/
[encoding]
```

The parameters are:

format:     xml or json
ID:         PubMed ID (such as 17299597) or PMC ID (such as PMC1790863)
encoding:   unicode or ascii

A sample URL:
```
https://www.ncbi.nlm.nih.gov/
research/bionlp/RESTful/pmcoa.cgi/BioC_xml/17299597/unicode
```
The same article in ASCII:
```
https://www.ncbi.nlm.nih.gov/
research/bionlp/RESTful/pmcoa.cgi/BioC_xml/17299597/ascii
```
Using JSON serialization instead of XML:
```
https://www.ncbi.nlm.nih.gov/
research/bionlp/RESTful/pmcoa.cgi/BioC_json/17299597/unicode
```
Using PMC ID instead of PubMed ID:
```
https://www.ncbi.nlm.nih.gov/
research/bionlp/RESTful/pmcoa.cgi/BioC_xml/PMC1790863/unicode
```

Note that PMC articles are available via both PubMed IDs and PMC IDs, whichever is more convenient for your project.

---

[6] https://github.com/ncbi-nlp/BioC-JSON

In addition to PMC, PubMed articles are also available via a RESTful webservice. The URLs are very similar to the PMC URLs, exactly as you would expect. The documentation is available at: https://www.ncbi.nlm.nih.gov/research/bionlp/APIs/BioC-PubMed/

## Results

At the time of this writing, there 1,907,370 articles in the PMC Open Access subset and 430,308 articles in the Author Manuscript Collection. With an overlap of 19,065 articles the combined collection is 2,318,613 articles. Of course, both of these collections are continuing to grow. PMC Open Access articles available via the API are updated daily. The Author Manuscript Collection is updated twice weekly. This is the same pace at which the PMC XML collections of these articles is updated. The FTP files are updated on a less frequent schedule. This is not a limitation because these files are primarily used for an initial creation of a local collection. These articles are all available in BioC XML and BioC JSON. They are also available in both Unicode (UTF-8) and ASCII encodings.

## Summary

More than 2.3 million PMC articles are available for text mining. This is a large, and growing proportion of the total. These articles are provided in both BioC XML and BioC JSON, formats that allow easy storing and sharing of annotations. They can be accessed via either a Web API or FTP. Unicode and ASCII encodings are both available. Convenient and growing access to full text biomedical research articles will lead to many exciting results.

## Funding

This research was supported by the Intramural Research Program of the NIH, National Library of Medicine.